\documentclass[10pt,aps,prx,showpacs,superscriptaddress,reprint,citeautoscript]{revtex4-1}

\usepackage{amssymb,amsmath,amsfonts,bm} 
\usepackage{graphicx}				\usepackage{color} 
\usepackage{dcolumn}   
\usepackage{bm}        
\usepackage{url}

\usepackage{hyperref}
\usepackage{subfigure}

\newtheorem{theorem}{Theorem}
\newtheorem{lemma}[theorem]{Lemma}

\def\eeq{\relax}
\def\beq#1#2\eeq{\begin{equation}\label{#1}#2\end{equation}}
\def\bal#1#2\eal{\begin{align}\label{#1}#2\end{align}}
\def\bse#1#2\ese{\begin{subequations}\label{#1}#2\end{subequations}}
\def\ba{\begin{aligned}}   \def\ea{\end{aligned}}

\def\Xint#1{\mathchoice
{\XXint\displaystyle\textstyle{#1}}%
{\XXint\textstyle\scriptstyle{#1}}%
{\XXint\scriptstyle\scriptscriptstyle{#1}}%
{\XXint\scriptscriptstyle\scriptscriptstyle{#1}}%
\!\int}
\def\XXint#1#2#3{{\setbox0=\hbox{$#1{#2#3}{\int}$}
\vcenter{\hbox{$#2#3$}}\kern-.5\wd0}}

\def\dashint{\Xint-}

\def\T{\tau}

\newcommand*{\CC}{{\mathbb{C}}}
\newcommand*{\RR}{{\mathbb{R}}}

\newcommand{\ii}{\ensuremath{\mathrm{i}}}
\newcommand{\E}{\ensuremath{\mathrm{e}}}

\DeclareMathOperator{\real}{Re}
\DeclareMathOperator{\imag}{Im}

\def\dd{\operatorname{d}} 

\def\rev#1{\textcolor{blue}{#1}}
\def\rev#1{{#1}}

\begin{document}

\title{Integral identities for   reflection, transmission and scattering coefficients
}
\author{Andrew N. Norris}
\affiliation{Mechanical and Aerospace Engineering, Rutgers University, Piscataway, NJ 08854-8058 (USA)}
\date{\today}

 \begin{abstract}

Several  integral identities related to acoustic  scattering are presented.  In each case  the identity involves the integral over  frequency of a physical quantity.  For instance, the integrated transmission loss, a measure of the transmitted acoustic energy through  an inhomogeneous layer, is shown to have a  simple expression in terms of spatially averaged physical quantities.
Known  identities for the extinction cross section and for the acoustic energy loss in a  slab with a rigid backing, are shown to be special cases of a general procedure  for finding  such integral identities.

 \end{abstract}

\maketitle

\section{Introduction}\label{sec1}

Identities for scattering coefficients that involve integrals over all frequencies, or equivalently over all wavelengths, provide a useful means to characterize scattering independent of frequency using a single parameter.  However, there are very few such identities available. 
An important example   is the 
integral of the extinction (which is the sum of the rate of energy absorption and  the scattering cross-section)
over all wavelengths \cite{Purcell69}. The {\it integrated extinction} (IE) is a 
 natural metric for quantifying   scattering reduction  \cite{Sohl2007,Monticone13}.  
The  IE has the important property that it is proportional to a linear combination of the monopole and dipole amplitudes if the scattering is {\it causal} \cite{Sohl07,Norris2015}, that is, the scattered wavefront  in the forward 
direction arrives after an equivalent plane wavefront in the background medium.    Causal scattering is the default for electromagnetics, although  there is no such limitation for acoustic or elastic waves.  
\rev{Many scattering situations  of interest in acoustics are non-causal, such as metal objects in water or air, for which the causal IE expression \cite{Purcell69,Sohl07} does not apply. However, by considering the scattering  in the time domain, it is possible to provide an expression applicable to all types of scatterers.   The generalization of Purcell's result to non-causal scattering can be found in \cite{Norris2015}. }

The only other integral identity known to the author relates the integral over all wavelengths of the acoustic absorption of a slab with rigid backing to the static effective bulk modulus of the slab 
\cite{Yang2017,Yang2017a}.  This result, based on work by Rozanov \cite{Rozanov2000} and on the Bode-Fano theorem \cite{Fano1950}, reduces  an integral of the logarithm of the absolute value of the acoustic  reflection  coefficient to a  form that can be interpreted in terms of static parameters plus a denumerable set of complex numbers defined by the zeros of the reflection coefficient as a function of frequency. 

The purpose of the paper is to present several new integral identities related to acoustic  scattering.  Some of these identities are similar to the one found previously \cite{Yang2017,Yang2017a}, requiring knowledge of the infinite set of zeros of a reflection coefficient.  However, new identities are presented which require only purely static physical  parameters, such as the total mass, or the effective compressibility. 

We begin in \S \ref{2=2} by considering some consequences of a signal being causal.  The acoustic scattering problem is defined in \S \ref{2.5}. The main results are given in \S \ref{3=3},  including integral identities for the reflection and transmission coefficients in one-dimensional  configurations. 

\section{Causal signal results} \label{2=2}
The real-valued signal is called {\it causal}  if it is zero before $t=0$,  
\beq{-1}
s(t) = 0, \ \ t<0   . 
\eeq
The   Fourier transform of the causal signal, 
\beq{-2}
S(\omega ) = \int_0^\infty s(t) e^{\ii\omega t} \dd t, 
\eeq
is  analytic in the upper half plane (or causal half plane) of the complex frequency $\imag \omega >0$. It may have zeros at the discrete set of frequencies $\{\omega_n \}$ in the upper half plane. 
 The additional property 
$S(-\omega^* ) = S^*(\omega ) $, with $^*$ the complex conjugate, follows from the fact that $s(t)$ is real. The low frequency  expansion of  the  Fourier transform  is 
\beq{-25}
S(\omega ) = S_0 + \ii \omega S_1 +   ( \ii \omega )^2 S_2 + {\cal O}(\omega^3) 
\eeq
where  $S_j$, $j=0,1,2,\ldots$ are 
 real valued.  The coefficients can be identified from \eqref{-2} as 
\beq{-2-1}
S_n = \frac 1{n ! }\int_0^\infty s(t) t^n \dd t .
\eeq
These integrals are well defined if the function $s(t)$ decays fast enough as $t\to \infty$, which is certainly true if the signal is of finite duration, as is  assumed here. 

The Fourier transform   of a causal function  satisfies the Sokhotski-Plemelj relations
for real values of $\omega$ \cite[eq.\ (1.6.7)]{Nuss72} 
\beq{6}
S(\omega )
 = \frac 1{\ii\pi } \dashint_{-\infty}^\infty 
\frac {S(\omega ')\dd \omega '}{\omega '- \omega }, 
\eeq
where $ \dashint$ denotes  principal value integral. Equation \eqref{6} is equivalent to $S(\omega ) = \ii \cal{H} (S)(\omega)$ where  $\cal{H}(S)$ is the Hilbert transform.   The real and imaginary parts of $S(\omega)$ on the real $\omega -$axis are therefore related to one another by the well known identities
$\imag S(\omega ) =  \cal{H} (\real S)(\omega)$ and 
$\real S(\omega ) =  -\cal{H} (\imag S)(\omega)$.  
The following identities  result  from expanding \eqref{6} about $\omega = 0$ for a real-valued signal, with details available in Appendix \ref{appA}, 
\bse{6--2}
\bal{6-2}
S_0
 &= \frac 2{\pi } \int_0^\infty  \imag S(\omega )
\frac {\dd \omega }{\omega   }, 
\\
S_1
 &= \frac 2{\pi } \int_0^\infty  \big( S_0 - \real S(\omega ) \big)
\frac {\dd \omega }{\omega^2   },  
\label{7b}
\\
S_2
 &= \frac 2{\pi } \int_0^\infty  \big( \omega S_1 -\imag S(\omega ) \big)
\frac { \dd \omega }{\omega^3   } , \ldots  .  
\eal
\ese

In dealing with acoustic transfer functions it is important to distinguish between  
 minimum phase and non-minimum phase  functions.
The canonical decomposition of a non-minimum phase transfer function is
\cite{Victor1989} 
\beq{7=2}
S(\omega ) = \E^{\ii \omega D} S_\text{mp} (\omega ) \prod_j Z(\omega , \omega_j)
\eeq
where $S_\text{mp}(\omega )$ is the unique minimum phase transfer function,
\beq{7=21}
  Z(\omega , u) = \frac {\omega -u}{\omega- u^*}  
\eeq
and  the set of complex frequencies 
$\{ \omega_j\}$ are in the causal half plane.
The delay $D$ is the largest value for which $s(t-D)$ is causal.  The minimum phase transfer function has no zeros in the upper half plane whereas $S(\omega ) $ has zeros at 
$\{ \omega_j\}$. \rev{Note that any zero of the form $\omega_j = \alpha_j +\ii
\beta_j$, $\omega_j \ge 0$, is accompanied by $\omega_j ' =- \alpha_j +\ii
\beta_j$.  This  ensures that  $S_\text{mp}(-\omega^* )= S_\text{mp}^*(\omega )$, and hence the causal time-domain function $s_\text{mp}^*(t)$ is     real-valued. }

Since $|S_\text{mp} (\omega )|=|S (\omega )|$ for real $\omega$, it follows that 
the real parts of the two functions $\ln S_\text{mp} (\omega )$ and $\ln S (\omega )$
coincide.  The imaginary parts of these two functions clearly differ, and most importantly, the real and imaginary parts of the minimum phase   function 
$\ln S_\text{mp} (\omega )$ are related by the Hilbert transform relations.  This property does not extend to the non-minimum phase  function.  

Minimum phase identification requires assumptions about the physical system \cite{McDaniel99,McDaniel2001}. If a transfer function, such as a reflection coefficient, is minimum phase then its phase as a function of frequency is uniquely defined by the amplitude. Conversely, the phase is not uniquely defined by the amplitude if the transfer function is not minimum phase. 

We next consider several  applications based on the low frequency behavior of minimum phase functions $S(\omega)$ with the common condition $S_0 = 0$.  The results all follow from the following identity, which is a consequence of \eqref{7b}.  
\begin{lemma}\label{1=1}
Let $S(\omega)$ be the Fourier transform of a causal real-valued signal with $S(0)=0$. 
Then 
\beq{789}
 \int_0^\infty   \real S(\omega ) 
\frac {\dd \omega }{\omega^2   }
 = -\frac {\pi }2 S_1
\eeq
where 
\beq{777}
S_1 = -\ii \left. \frac{\dd S}{\dd \omega} \right|_{\omega = 0} .
\eeq
\end{lemma}

\rev{The results in \S  \ref{2=2}  are based upon a causal scattering process; that is, the forward scattered signal follows the incoming signal.  In the absence of material damping when the wave speed is real, there is no ambiguity in the meaning of causal.  With absorption present, the strict definition requires considering how a sharp delta pulse transmits.    }

\section{Acoustic scattering }\label{2.5}

The  acoustic   pressure $p({\bf x}) \in \CC$ satisfies the  Helmholtz equation 
outside of a finite region $\Omega$, the scatterer, 
 \beq{1}\nabla^2 p + k^2 p = 0, \ \  {\bf x}\in \RR^d /\Omega . 
\eeq
The system may be one, two or three-dimensional, $d=1,2$ or $3$.  
Time harmonic dependence is considered with  $k=\omega /c$ and $c$ is the sound speed,  $c = (C\rho)^{-1/2}$ 
 where  the uniform exterior acoustic medium has mass density $\rho$   and compressibility $C$.  The   factor  $e^{-i\omega t}$ is understood and omitted.

The scatterer, $\Omega$, may be an inhomogeneous acoustic or elastic object.  The specific results will be limited to acoustic scatterers of density $\rho ' $ and compressibility $C'$. 
Damping in the scatterer may be included by considering the material properties as frequency dependent complex parameters  $\rho ' ({\bf x}, \omega) $,  $C'({\bf x}, \omega)$ and derived quantities, the {wave speed} $c'({\bf x}, \omega) = (C '\rho ')^{-1/2}$ and impedance $z'({\bf x}, \omega) = ( \rho '/C' )^{1/2}$.   The zero frequency limits, or static values,  will play an important role in our results, and we therefore denote them 
\beq{398}
\begin{aligned}
\rho_0 ' &=\rho ' ({\bf x},0) , \ \ 
C_0 ' =C ' ({\bf x},0) ,  \\ 
c_0 ' &=c ' ({\bf x},0) ,  \ \ 
z_0 ' =z ' ({\bf x},0) .
\end{aligned}
\eeq 
Note that these are necessarily real-valued quantities. 


The total  acoustic pressure $p$ comprises  an incident plane wave $e^{\ii k x}$ 
plus  the   scattered pressure $p_s$, 
\beq{=2}
p =  e^{\ii k x} + p_s ({\bf x}) . 
\eeq
The  scattering amplitude $F(\theta ,\omega)$ is defined by 
\beq{3}
p_s  =    F(\theta ,\omega) \, 
\Big( \frac k{\ii 2\pi r}\Big)
^\frac{d-1}2 \, e^{\ii k r}\, \big[ 1 + 
{\cal O}\big(\frac 1{kr}\big) \big]  
\eeq
as $r=| {\bf x}| \to \infty $, where $\theta$ is the scattering direction,  
 $\theta = 0$ corresponding to the direction of incidence $\hat{\bf k}$. 
Note that \eqref{3} is exact in one dimension, in which case $\theta$ only takes the values $0$ and $\pi$, with 
\beq{3-1}
T(\omega) = 1+  F(0 ,\omega), \ \ 
R(\omega) =  F(\pi ,\omega) 
\eeq
the transmission and reflection coefficients, respectively. 

\section{Integral identities }\label{3=3}

\subsection{Integrated extinction }\label{4-9}

The  {\it extinction} cross section   is defined as $\sigma = \sigma_\text{sc} + \sigma_\text{ab}$
with $\sigma_\text{ab}$  the absorption cross section (zero in the absence of loss),  and $\sigma_\text{sc} $ is the  scattering cross section 
\beq{4}
\sigma_\text{sc} (\omega)=  \int |p_s|^2 \dd s 
\eeq
where the integral is around any  surface enclosing the scatterer.   
The  { optical theorem}   relates the extinction to  the forward scattering amplitude,   
\beq{5}
 \sigma  = - 2\, \real S(\omega)  \ \ \text{where} \  S(\omega) \equiv F(0,\omega)  .
\eeq
The  {\em integrated extinction} (IE),  
\beq{27}
\int_0^\infty 
\frac {\sigma(\omega )}{\omega^2}\, \dd \omega  \, \ge 0,
\eeq
defines the  total cross section  over all frequencies. 

It follows from Lemma \ref{1=1}, Eqs.\  
\eqref{5},  \eqref{27} and the fact that the forward scattering amplitude vanishes at zero frequency ($S_0=0$) that 
 \beq{7}
\int_0^\infty \frac {\sigma(\omega )}{\omega^2}\, \dd \omega 
= \pi S_1 . 
\eeq
The identity  \eqref{7}  for $d=3$ was derived by Purcell \cite{Purcell69} for electromagnetics and was first used in acoustics by   Sohl et al.\ \cite{Sohl07}.  
Equation \eqref{7} is, however, restricted to scattering for which the forward scattered impulse function (the time domain version of $S(\omega)$) is strictly  causal.  This is always the case if the wave speed in the scatterer is everywhere less than that of the exterior medium.  However, if the scatterer comprises faster material such that the forward amplitude precedes the direct wave in time, then  the function $S(\omega )$ is no longer analytic in the upper half plane, and \eqref{7} is not valid.   The problem arises from the strict definition of the scattered amplitude in Eq.\ \eqref{3} which allows use of the optical theorem.    Resolution of this issue can be found in \cite{Norris2015} which describes the generalization of \eqref{7} to all possible scatterers.  Here we will only consider scattering such that \eqref{7} holds.

The zero frequency limit in \eqref{7} allows us to interpret $S_1$ and hence the IE   in terms of quasistatic properties.  For instance, if the 
scatterer has  volume $|\Omega|$, compressibility  $C ' ({\bf x},\omega)$ and uniform density  $\rho '(\omega)$, then \cite{Sohl07} 
\beq{-2723}
\int_0^\infty \frac {\sigma(\omega )}{\omega^2}\, \dd \omega 
= \frac {\pi}{2c} \Big( 
	\big( \frac{\langle C_0'\rangle}C -1 \big) |\Omega| - \hat{\bf k}\cdot  
	{\boldsymbol \gamma}\Big(\frac{\rho}{\rho_0'} \Big)\cdot \hat{\bf k} \Big)  
\eeq
where ${\boldsymbol \gamma}$ is the polarizability dyadic \cite{Dassios00}  proportional to  $|\Omega|$ and 
$\langle\cdot  \rangle $ is  the spatial average, e.g. 
\beq{-45}
 \langle C_0'\rangle = \frac 1{|\Omega|} \int_\Omega C_0' ({\bf x}) \dd {\bf x}. 
\eeq

\rev{The compressibility term in Eq.\ \eqref{-2723} is the monopole contribution to the scattering, which is independent of the direction of observation.  The polarizability produces a dipole field with dependence $\hat{\bf k}\cdot  
	{\boldsymbol \gamma} \cdot \hat{\bf x} $ where $ \hat{\bf x}$ is the unit vector in the scattering direction.}
The identity for the IE is therefore a special case of the more general integral equality
\beq{-2=3}
- 2\, \real 
\int_0^\infty \frac {F(\theta,\omega )}{\omega^2}\, \dd \omega 
= \frac {\pi}{2c} \Big( 
	\big( \frac{\langle C_0'\rangle}C -1 \big) |\Omega| - \hat{\bf k}\cdot  
	{\boldsymbol \gamma}\Big(\frac{\rho}{\rho_0'} \Big)\cdot \hat{\bf x} \Big) .
\eeq
  For instance, if the scatterer is a uniform sphere with sound speed $c'\le c$
	\cite[p.\ 282]{Rayleigh1878} 
\beq{-2=-3}
- 2\, \real 
\int_0^\infty \frac {F(\theta,\omega )}{\omega^2}\, \dd \omega 
= \frac {\pi |\Omega|}{2c} \Big( 
	\frac{ C_0'}C -1  + \frac{3 (\frac{\rho_0 '}\rho  - 1)}{2\frac{\rho_0 '}\rho + 1} \cos\theta 	
	\Big)  
\eeq
where $\cos\theta =\hat{\bf k}\cdot   \hat{\bf x} $.
The integral \eqref{-2=-3} is  positive for $\theta = 0$ but may be negative for other directions.

\subsection{Transmission and reflection    from a slab }
Consider a 1D system with non-uniform density and compressibility  $ \rho '(x,\omega)$, $C'(x,\omega)$  restricted to $\Omega = [0,a]$. 
The reflection and transmission coefficients are given by Eq.\ \eqref{9-5}.
Lemma \ref{1=1} with $S(\omega )=R(\omega )$ from \eqref{9-5} implies the identity
\beq{-32}
\int_0^\infty  \real R(\omega )
\frac {\dd \omega }{\omega^2   }
= \frac{\pi a}{4c} \Big( \frac{\langle \rho_0' \rangle}\rho - \frac{\langle C_0'  \rangle} C  \Big) .
\eeq

The choice  $S(\omega )=T(\omega )$ is not useful since $T(0)$ is non-zero.  An alternative is to consider $S(\omega )=T(\omega )-1$, which reproduces 
Eq.\ \eqref{-2723} for 1D wave propagation.  In this case, the IE reduces to \cite{Norris2015} 
\beq{10}
\int_0^\infty \frac {\sigma(\omega )}{\omega^2}\, \dd \omega 
= \frac{\pi a}{2c}
\Big( 
 \frac{ \big\langle C_0'  \rangle}C +\frac{\langle \rho_0' \rangle}\rho - 2 \Big).
\eeq
Again we note that this formula is only valid if the travel time across the slab is less than that in a slab of the same width of the external  fluid; i.e.  the forward scattering is causal.
 The extension of Eq.\ \eqref{10} to the non-causal situation  is discussed in \cite{Norris2015}.

Another option is to consider $S(\omega )=\ln T(\omega )$ which has $S_0= 0$, and we can therefore use Lemma \ref{1=1}, with careful consideration for the fact that the parameter $S_1$ is that for the minimum phase function $\ln T_\text{mp}(\omega )$.  \rev{The transmission $ T(\omega )$ does not have zeros in the upper half plane, as can be seen from Eq.\ \eqref{9-3}, and hence 
$\{ \omega_j\} = \O $, see Eq.\ \eqref{7=2}.} Therefore, the minimum phase transmission coefficient is defined by the earliest time at which the transmitted impulse response becomes non-zero.  This depends upon the difference in travel time through the slab and through the same fluid distance.  Thus,   
\beq{93=}
T_\text{mp}(\omega )=
T(\omega ) \E^{\ii \omega (\T - \T ')}
\eeq
where  $\T = \frac ac $ is the travel time across  an equivalent slab of fluid, and $\T '$ is the travel time across the slab, defined below. 
Note that the real part of the logarithm of $T(\omega )$ and $T_\text{mp}(\omega )$ are the same for real valued $\omega$.

Taking $S(\omega ) = \ln T_\text{mp}(\omega )$ and noting (i)   $S(0)=0$,  (ii)   $\real S(\omega ) = \ln |T (\omega )|$, and (iii) that the low frequency expansion of $S$ is $S(\omega ) = \ii \omega T_{\text{mp},1} +\ldots$, 
we may use  Lemma \ref{1=1}  in the form 
\beq{7=32}
- \int_0^\infty \ln |T (\omega )|
\frac {\dd \omega }{\omega^2   } = \frac {\pi }2
T_{\text{mp},1} .  
\eeq
The coefficient $T_{\text{mp},1}$ in turn follows from \eqref{93=} and 
\eqref{9-5}, to give 
\beq{-325}
-\int_0^\infty  \ln |T(\omega )|^2
\frac {\dd \omega }{\omega^2   }
= \frac{\pi a}{2c} \Big(  \frac{\langle C_0'  \rangle} C  
+ \frac{\langle \rho_0' \rangle}\rho - 2 \frac{\T'}{\T}\Big) .
\eeq
This quantity represents the total transmitted energy loss over all frequency, and we therefore call it the {\it integrated transmission loss} (ITL). 

 {In order to further simplify Eq.\ \eqref{-325} we first consider the slab with no  absorption.  The wave speed $c'$ and impedance $z'$ are then independent of frequency, yielding }
\beq{698}
 \T '  =  \int_0^a \frac {\dd x }{c'(x)}  .
\eeq
The integrated transmission loss   can then be expressed in a form that is clearly non-negative
\beq{-326=}
-\int_0^\infty  \ln |T(\omega )|^2
\frac {\dd \omega }{\omega^2   }
= \frac{\pi a}{2} 
\Big\langle  \frac 1{c'}
\Big( \sqrt{\frac{z'}z} - \sqrt{\frac z{z'}}  \Big)^2 
\Big\rangle .
\eeq

{The presence of absorption implies a wave speed in the slab that is frequency dependent: $c'(x,\omega)$.   The travel time  $\T '$ should then be understood as the time taken for the first arrival of a sharp pulse, which is defined by the infinite frequency limit 
\beq{-4=4}
c_\infty ' (x) = \lim_{\omega \to \infty} c'(x,\omega).
\eeq
This is real-valued satisfying $c_\infty ' \ge c_0'$.  The travel time $\T '$ is therefore
\beq{69-8}
 \T '  =  \int_0^a \frac {\dd x }{c_\infty'(x)}  .
\eeq
The  general version of the identity \eqref{-326=} that includes absorption is  
\beq{-32=}
-\int_0^\infty  \ln |T(\omega )|^2
\frac {\dd \omega }{\omega^2   }
= {\pi a} 
\Big\langle  \frac 1{2c_0'}
\Big( \sqrt{\frac{z_0'}z} - \sqrt{\frac z{z_0'}}  \Big)^2 
+ \frac 1{c_0'}- \frac 1{c_\infty '}
\Big\rangle  
\eeq
where, as usual, $c_0' (x) = c'(x,0)$ and $z_0' (x) = z'(x,0)$
The property  $c_\infty ' \ge c_0'$,  guarantees a non-negative ITL. }

\subsubsection{Numerical example of attenuated transmission}

We consider a standard linear solid model, also known as Zener's model \cite{Zener}, for the slab bulk modulus.  The stress, $\sigma$ $(= -p)$, and dilatational strain $\varepsilon$, are related by 
$\sigma + \eta \partial_t \sigma = K_0(
\varepsilon + \eta_1 \partial_t \varepsilon )$, with $\eta_1 > \eta > 0$. The effective bulk modulus is then $K ' (\omega ) = K_\infty \big( 1 - \alpha /(1-\ii \omega \eta)\big)$
where $K_\infty = K_0 \eta_1 / \eta $, $\alpha = 1-\eta  / \eta_1 $. The acoustic speed 
is $c' = \sqrt{K'/\rho '}$, or 
\beq{4=7}
c '(\omega )  = c_\infty ' \Big( 1 - \frac{\alpha}{1-\ii \omega \eta}\Big)^{1/2} 
\eeq
where  $c_\infty ' = \sqrt{K_\infty/\rho '}$.  Hence, $c_0 ' = c_\infty ' (1-\alpha)^{1/2}$. 

\begin{figure}[htbp]
				\begin{center}	
				\includegraphics[width=3.6in , height=2.6in 					]{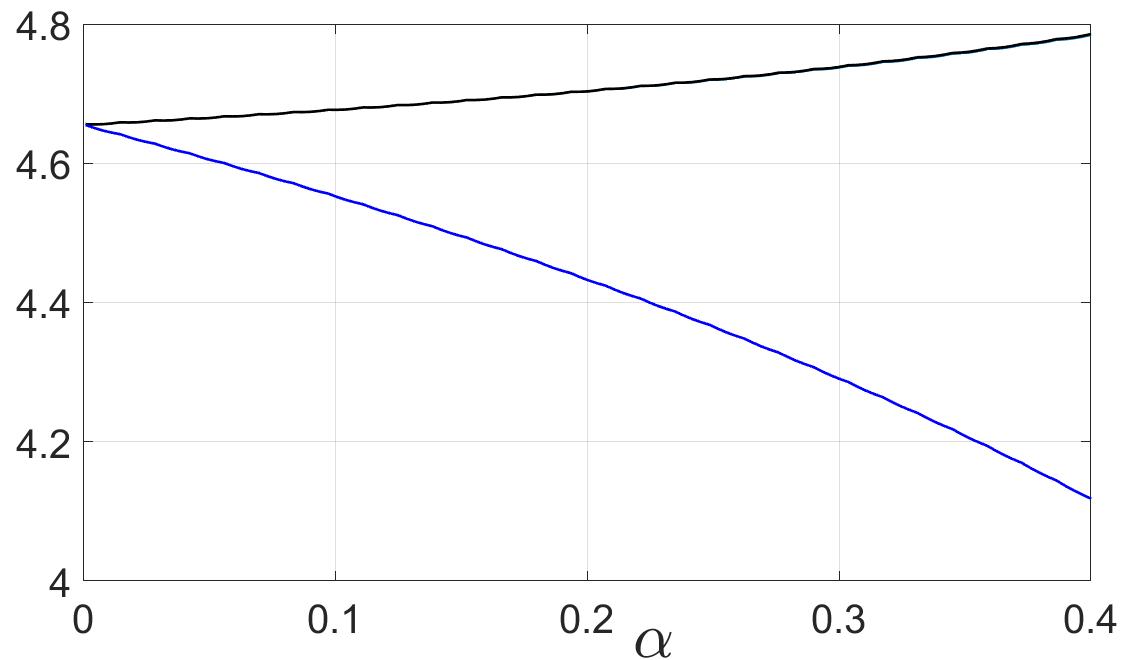} 
	\caption{ The integral 
$-\int_0^\infty  \ln |T(\omega )|^2 \omega^{-2 } 
 {\dd \omega } $  for the attenuation  model of Eq.\ \eqref{4=7}.  The upper (black) curve is actually two close to one another, one  determined using numerical integration, the other from the expression in Eq.\ \eqref{-32=}.  The lower (blue) curves use the  expression in Eq.\ \eqref{-326=} which does not account for damping.}
		\label{fig1} \end{center}  
	\end{figure}

In the numerical example the background medium properties are $c=\rho =  1$. The slab properties are $a=1$, $\rho ' =4.3$, 
$c_\infty ' = 1.37$ and $\eta = a/(10\pi c_\infty ' )$.  We consider values of $\alpha $ from zero (no damping) to $\alpha =0.4$. Figure \ref{fig1} shows three curves: (i) the integral 
$-\int_0^\infty  \ln |T(\omega )|^2 \omega^{-2 } 
 {\dd \omega } $ evaluated numerically, (ii) the expression on the right side of Eq.\ 
\eqref{-32=}, and (iii) the expression on the right side of Eq.\ 
\eqref{-326=}.  The curves (i) and (ii) are coincident within the accuracy of the (crude) numerical integration scheme. Curve (iii), which is only valid for the lossless case, agrees  with the others in that limit but diverges from them as the damping grows. 

\subsection{Reflection    from a slab with rigid or free backing}

\rev{A  reflected delta pulse signal $s(t) = \delta (t)$ has zero delay because of the instantaneous wavefront interaction at the interface.  The associated  minimum phase function therefore has no phase delay but it does involve an all-pass filter associated with the zeros $\omega_j$ of $R(\omega )$ in the upper half plane, 
\beq{3=11}
R_\text{mp}(\omega ) =
R(\omega ) / \prod_j Z(\omega , \omega_j) .
\eeq
Let  $R(\omega )= R_0 + \ii \omega R_1 +\ldots$, then 
\beq{3=16-}
R_\text{mp}(\omega ) =R_{\text{mp},0} + \ii \omega R_{\text{mp},1} +
\ldots , 
\eeq
where   
\beq{3=16}
\begin{aligned}
R_{\text{mp},0} & = Z_0 R_0,
\\
  R_{\text{mp},1} &= Z_0 \Big(R_1 + 2 R_0 \sum_j \imag \frac 1{\omega_j} \Big)
	\end{aligned}
\eeq
and $Z_0 = \prod_n (\omega_n^*/\omega_n )$.  The constraints on  $\{ \omega_n\}$ imply that $Z_0 = 1$ or $Z_0 = -1$.    In the next examples we consider the limiting cases for which $R_0 =\pm 1$.  
}

\subsubsection{Rigid backing}
For the rigid backing, Lemma \ref{1=1} and Eqs.\ \eqref{3=16}, \eqref{9-52} imply using 
\rev{$S(\omega ) = \ln \big(Z_0 R_\text{mp}(\omega ) \big)$} that 
\beq{-326}
-\int_0^\infty  \ln |R(\omega )|^2
\frac {\dd \omega }{\omega^2   }
= \frac{2\pi a}{c}    \frac{\langle C_0'  \rangle} C  
+ 2\pi \sum_j \imag \frac 1{\omega_j}    
\eeq
in agreement with \cite[Eq.\ (A9)]{Yang2017} and 
\cite[Eq.\ (S9)]{Yang2017a}. 

Alternatively, taking $S(\omega ) =  R (\omega ) -1$, Lemma \ref{1=1} and Eq.\  \eqref{9-52} yield
\beq{5-88}
\int_0^\infty  \real \big(1- R(\omega ) \big)
\frac {\dd \omega }{\omega^2   }
= \frac{\pi a}{c}    \frac{\langle C_0'  \rangle} C .
\eeq

\subsubsection{Soft backing}
For the soft backing, Lemma \ref{1=1} and Eqs.\  \eqref{3=16}, \eqref{9-53} imply using 
\rev{$S(\omega ) = \ln [-Z_0 R_\text{mp}(\omega )]$} that 
\beq{-345}
-\int_0^\infty  \ln |R(\omega )|^2
\frac {\dd \omega }{\omega^2   }
= \frac{2\pi a}{c}    \frac{\langle \rho_0 '  \rangle} \rho  
+ 2\pi \sum_j \imag \frac 1{\omega_j}   . 
\eeq
Alternatively, taking $S(\omega ) =  R (\omega ) +1$, Lemma \ref{1=1} and Eq.\  \eqref{9-53} yield
\beq{5-58}
\int_0^\infty  \real \big(1+ R(\omega ) \big)
\frac {\dd \omega }{\omega^2   }
= \frac{\pi a}{c}    \frac{\langle \rho_0 '  \rangle} \rho  .
\eeq

\section{Discussion: Connection with the mass law} \label{5=5}

The main results are new  integral identities \eqref{-32} and \eqref{-326=} for the reflection and transmission coefficients of  a slab in an infinite medium, and  Eqs.\  \eqref{5-88} -  \eqref{5-58} for reflection from a slab with a rigid or soft backing.   The general methodology has also been used to  derive two previously known identities:  \eqref{-2723} for the integrated extinction and \eqref{-326} for the slab with rigid backing. 


It is important to point out that all of these results include the possibility of energy loss through material damping.  {However, many of the  integral identities depend only on the limiting static values of the density and bulk modulus, e.g. Eq.\  \eqref{-2723} for the IE, and Eqs.\ \eqref{-32}, \eqref{5-88}, \eqref{5-58} for reflection coefficients.  These identities are therefore independent of the particular damping mechanisms present,   an unexpected and surprising result. 
The identity \eqref{-32=} for the integrated transmission loss depends not only on the static values of the slab parameters but also  on the  infinite frequency value of the wave speed, which cannot be less than the zero frequency speed.
}

The identities \eqref{-326} and \eqref{-345}  involve the complex-valued zeros $\{\omega_j\}$ which do depend on the material damping.  
In the absence of absorption, since the slab is backed by a perfect reflector it follows that $|R(\omega)|=1$ in both cases.  The integrals \eqref{-326} and \eqref{-345} are therefore zero, with the right hand sides implying two identities  for the quantities $\sum_j \imag \frac 1{\omega_j}$.   
When  damping is present the integrals represent the  loss of acoustic energy into the slab over all frequencies.  This was the motivation for the original derivation  \cite{Yang2017,Yang2017a} of \eqref{-326}.

Finally we note an interesting connection between the exact identity  \eqref{-326=} for the  integrated transmission loss of  a uniform slab with no damping, 
\beq{4-3}
-\int_0^\infty  \ln |T(\omega )|^2
\frac {\dd \omega }{\omega^2   }
= \frac{\pi a \rho '}{2c\rho} 
\Big( 1 - \frac z{z'}  \Big)^2  
\eeq
and the same integral using a well known and useful approximation for $T(\omega)$.  The transmission coefficient using the "mass law"  \cite[\S 6.7]{KinslerFrey} is 
\beq{8-2}
T_\text{mass}(\omega) = \frac 1{1- \ii \frac{\omega a \rho'}{2 c \rho}} .
\eeq
This yields an integrated transmission loss 
\beq{4-4}
-\int_0^\infty  \ln |T_\text{mass}(\omega )|^2
\frac {\dd \omega }{\omega^2   }
= \frac{\pi a \rho '}{2c\rho} 
\eeq
clearly a good approximation to the exact ITL \eqref{4-3} if $z'  \gg z$, which is implicitly assumed in the mass law approximation.  It is interesting to note that the simple mass law approximation captures the full frequency content of the integrated transmission loss. 

This all suggests a slight modification of the mass law, 
\beq{8-9}
T_\text{approx}(\omega) = \frac 1{1- \ii \frac{\omega a \rho'}{2 c \rho}
\big( 1 - \frac z{z'}  \big)^2 
} . 
\eeq
The proposed transmission coefficient has several benefits including that it is unity if the impedances are equal, as it should.  It also  reproduces the integrated   loss \eqref{4-3} exactly.    However, the mass law  in its simple or modified form 
cannot be expected to accurately reproduce  the ITL for a slab with absorption, 
Eq.\ \eqref{-32=}, since the approximations \eqref{8-2} and \eqref{8-9} for the transmission coefficient use static quantities only. 

\bigskip
\noindent\textbf{Acknowledgments}

\noindent  Thanks to Allan P. Rosenberg and to the reviewers for comments. This work  was supported under the National Science
Foundation  Award No. EFRI 1641078  and the ONR MURI Grant No. N000141310631.

\appendix

\section{Integral identities for real causal signals}\label{appA}

The $\omega-$functions $S(\omega)$, $1/(\omega - \zeta)$ are both transforms of causal signals, and therefore so are $S(\omega)/(\omega - \zeta)$ and the {\it subtraction} function 
\cite[eq.\ (1.7.4)]{Nuss72} 
\beq{6-10}
S^{(1)}(\omega , \zeta) = 
\frac{ S(\omega) - S(\zeta)}{\omega - \zeta}. 
\eeq
In this way we may form a chain of causal transforms: $S^{(n)}(\omega)$, $n=1,2,\ldots$
\beq{6-11}
S^{(n+1)}(\omega , \zeta) = 
\frac{ S^{(n)}(\omega , \zeta) - {\dd^{n} } S(\omega) /{\dd \omega^{n } } 
}{\omega - \zeta} , \ \ n >1 .
\eeq
Being  causal transforms, the Sokhotski-Plemelj relation \eqref{6}  applies to each of the functions $S^{(n)}$, 
\beq{6=1}
S^{(n)}(\omega , \zeta)
=
\frac 1{\ii \pi}
\dashint_{-\infty}^\infty S^{(n)}(\omega ', \zeta)
\frac{\dd \omega ' }{\omega '- \omega }.
\eeq
The limiting values of the $S^{(n)}$ functions as $\zeta \to \omega$ are 
\beq{6=12}
S^{(n)}(\omega , \omega) = \frac 1{n!}\frac{\dd^{n} S (\omega)} {\dd \omega^{n } } 
\eeq
from which it follows  that the $n-$th derivative of $S(\omega)$ can be expressed as an integral of lower order derivatives, 
\beq{6=13}
\frac{\dd^{n} S (\omega)} {\dd \omega^{n } } 
=
\frac {n!}{\ii \pi}
\dashint_{-\infty}^\infty S^{(n)}(\omega ', \omega)
\frac{\dd \omega ' }{\omega '- \omega }.      
\eeq
Specializing the Sokhotski-Plemelj relation \eqref{6} and the identities \eqref{6=13} 
to the case $\omega = 0$ yields 
\bse{6--1}
\bal{6-1}
S( 0 )
 &= \frac 1{\ii\pi } \dashint_{-\infty}^\infty 
S(\omega )
\frac {\dd \omega }{\omega   }, 
\\
S'( 0 )
 &= \frac 1{\ii\pi } \dashint_{-\infty}^\infty 
\big( S(\omega ) - S(0) \big) 
\frac { \dd \omega}{\omega^2   } , \label{6-19}
\\
S''( 0 ) 
&=
{ \frac 2{\ii\pi }} \dashint_{-\infty}^\infty 
\big( S(\omega ) - S(0) - \omega S'( 0 ) \big) 
\frac { \dd \omega}{\omega^3   } \ldots   .  \label{6-12}
\eal
\ese

Finally, we restrict \eqref{6--1} to {\it real causal} signals for which we have the additional property $S(-\omega ) = S^*(\omega ) $. Using this  and \eqref{-25} the integrals \eqref{6--1}  reduce to 
\eqref{6--2} .

\section{Layered one dimensional medium}\label{appB}
 
The slab occupies $\Omega: x\in [0,a]$ with non-uniform density and compressibility  $ \rho '(x,\omega)$, $C'(x,\omega)$.  
The 2-vector of particle velocity and acoustic pressure ${\bf U} = \big( v, p\big)^T$ is propagated from one end to the other by the 2$\times$2  matrix $ {\bf M}(x,\omega)$, $\det {\bf M} = 1$, such that ${\bf U}(a) ={\bf M}(a,\omega) {\bf U} (0)$. 
The propagator satisfies \cite[\S 7]{Pease}
\beq{8}
\frac {\dd{\bf M} }{\dd x} (x,\omega) = i\omega {\bf Q}{\bf M}, \ \ 
{\bf M}(0,\omega) = {\bf I}, 
\eeq
with ${\bf I}$  the identity and 
\beq{-56}
 {\bf Q}(x,\omega) = 
\begin{pmatrix} 0 & C'(x,\omega)  \\ \rho ' (x,\omega)& 0 
\end{pmatrix} .
\eeq
The solution   follows using well known methods for uni-dimensional systems, e.g. Ch.\ 7 of Pease \cite{Pease}.  The medium in $x<0$  ($x>a$) is assumed to have properties $z,c$ ($z_1,c$), where 
the impedance $z_1$ is introduced to allow for different boundary conditions at $x=a$, specifically the cases 
of interest $z_1=z, \infty, 0$. 

The reflected and transmitted  fields  are   
\beq{-34}
p(x) = \begin{cases}
e^{\ii k x} + R e^{-\ii k x} , & x<0 ,
\\
T e^{\ii k x} , & x>a ,
 \end{cases}
\eeq
where $k=\omega /c$.   Hence, 
\bse{0-3}
\bal{9-3}
T &= \frac{2 \E^{-\ii ka} }{ M_{11} + zz_1^{-1} M_{22}-z M_{12} - z_1^{-1} M_{21}},
\\
R &= \frac{ M_{11} - zz_1^{-1} M_{22}+z M_{12} - z_1^{-1} M_{21}}{ M_{11} + zz_1^{-1} M_{22}-z M_{12} - z_1^{-1} M_{21}},
\eal
\ese
where $M_{ij}$ are the elements of  $ {\bf M}(a,\omega)$.

For our purposes, we note that 
 at low frequency  ${\bf M}(a,\omega) = {\bf I} + i\omega a \langle {\bf Q}_0 \rangle + ...$ where $\langle \cdot \rangle $ denotes the average value in  $\Omega$.  
Hence,
\beq{0-4}
\ba
T &=  \frac{ 2z_1 }{ z_1 +z} 
+\frac{ \ii k a 2z_1^2}{ (z_1 +z)^2} 
\big( \frac{ \big\langle C_0'  \rangle}C + \frac{z\langle \rho_0' \rangle}{z_1\rho}
- \frac z{z_1}-1 \big) 
 \\ &
+ {\cal O}(\omega^2 ), 
\\
R &= \frac{ z_1 -z}{ z_1 +z} +\frac{ \ii k a 2z_1^2}{ (z_1 +z)^2} 
\big( \frac{ \big\langle C_0'  \rangle}C - \frac{z^2\langle \rho_0' \rangle}{z_1^2\rho}
  \big) 
+ {\cal O}(\omega^2 ).
\ea
\eeq
The three cases of interest are (i) the slab sandwiched by the same material on either side, (ii) the slab with a rigid backing, and (iii) the slab with a soft boundary on one side, or respectively
\bse{0-5}
\bal{9-5}
&\left.\ba
T &=  1 +  \frac{ \ii k a}{2 } 
\Big( 
 \frac{ \big\langle C_0'  \rangle}C +\frac{\langle \rho_0' \rangle}\rho - 2 \Big)
+ {\cal O}(\omega^2 ),
\\
R &= \frac{ \ii k a}{2 } 
\Big( 
 \frac{ \big\langle C_0'  \rangle}C -\frac{\langle \rho_0' \rangle}\rho  \Big)
+ {\cal O}(\omega^2 ), 
\ea \right\}
\ \ z_1 =z, 
\\
&R = 1 +  \ii k  a 2 \frac{ \langle C_0' \rangle }{C} 
+ {\cal O}(\omega^2 ), \ \ z_1 = \infty , \label{9-52}
\\
&R= -1 -\ii k   a 2  \frac{\langle \rho_0 ' \rangle  }{\rho}
+ {\cal O}(\omega^2 ), \ \ z_1 = 0.  \label{9-53}
\eal
\ese

Finally, note that damping may be included by considering the material properties as frequency dependent complex parameters.  In that case $\langle \cdot \rangle$ is the spatial average of the real valued static quantity $(\omega = 0)$.


\end{document}